\documentclass[12pt,a4paper,twoside]{article}

\usepackage{amsfonts,amsmath,amssymb,amsthm,latexsym}

\theoremstyle{definition}

\theoremstyle{remark}

\def\1{{\bf 1}}
\def\id{\mbox{id\,}}
\def\ot{\!\otimes\!}

\def\F{\mbox{$\cal F$}}

\def\bF{\mbox{$\overline{\cal F}$}}

\def\f{{\scriptscriptstyle {\cal F}}}

\def\cross{\mbox{$\rule{0.7pt}{1.3ex}\!\times $}}

\def\ra{\rangle}
\def\la{\langle}
\def \A{{\cal A}}
\def \B {{\cal B}}

\def \D {{\cal D}}
\def \E {{\cal E}}

\def \H {{\cal H}}
\def \bH {\overline{\cal H}}

\def \O {{\cal O}}

\def \P {{\cal P}}

\def \L {{\cal L} }

\def \X {{\cal X} }
\def \Hau{{\sf H}^{\scriptscriptstyle (1)}}

\def \Haus{{\sf H}^{\scriptscriptstyle (1)}_{\scriptscriptstyle \star}}
\def \Hans{{\sf H}^{\scriptscriptstyle (n)}_{\scriptscriptstyle \star}}
\def \Has{{\sf H}_{\scriptscriptstyle \star}}
\def \varphihs{\varphi^{\scriptscriptstyle H}_{\scriptscriptstyle \star}}

\def \psin{\psi^{\scriptscriptstyle (n)}}

\def\R{\mbox{$\cal R$}}

\def\hA{\mbox{$\widehat{\cal A}$}}

\def\Ha{{\sf H}}
\def\hH{\mbox{$\hat{H}$}}

\newcommand{\trc}{\triangleright}
\def\g{\mbox{\bf g\,}}

\def\b#1{{\mathbb #1}}
\def\nn{\nonumber \\}

\newcommand{\be}{\begin{equation}}
\newcommand{\ee}{\end{equation}}
\newcommand{\bea}{\begin{eqnarray}}
\newcommand{\eea}{\end{eqnarray}}
\newcommand{\ba}{\begin{array}}
\newcommand{\ea}{\end{array}}
\renewcommand{\thefootnote}{\fnsymbol{footnote}}
\title{\MakeUppercase{Noncommutative spaces with twisted
symmetries and second quantization}}
\author{Gaetano Fiore$^{1,2}$}
\date{}
\renewcommand{\date}{\vspace{-5mm}}
\begin{document}
\maketitle \vspace*{-3mm}\relax
\renewcommand{\thefootnote}{\arabic{footnote}}

\noindent \textit{\small $^1$Dip. Matematica e Applicazioni, Universit\`a ``Federico II'',
   V. Claudio 21, 80125 Napoli, Italy}\\
\noindent \textit{\small $^2$ I.N.F.N., Sez. di Napoli,
        Complesso MSA, V. Cintia, 80126 Napoli, Italy}

\begin{abstract}
\noindent In a minimalistic view, the use of noncommutative coordinates
can be seen just as a way to better express non-local interactions of a
special kind:
1-particle solutions (wavefunctions) of the equation of motion in the
presence of an external field may look simpler as functions of noncommutative
coordinates. It turns out that also the wave-mechanical description of a
system of $n$ such bosons/fermions and its second quantization is simplified
if we translate them in terms of their deformed counterparts. The latter are
obtained by a general twist-induced
$\star$-deformation procedure which deforms in a coordinated way not just the
spacetime algebra, but the larger algebra generated by any number $n$ of
copies of the spacetime coordinates and by the particle creation and
annihilation operators. On the deformed algebra the action of the
original spacetime transformations looks twisted.

In a non-conservative view, we thus obtain a twisted covariant framework for QFT
on the corresponding noncommutative spacetime consistent with quantum
mechanical axioms and Bose-Fermi statistics. One distinguishing
feature is that the field commutation relations remain of the type
``field (anti)commutator=a distribution''. We illustrate the results
by choosing as examples interacting non-relativistic and free relativistic
QFT  on Moyal space(time)s.
\end{abstract}

\section{Introduction}

Second Quantization played a crucial role in the foundation of
Quantum Field Theory (QFT) as a bottom-up approach from the
wave-mechanical description of a system of $n$ identical quantum
particles. The nonrelativistic field
operator of a spinless particle in $\b{R}^3$
(in the Schr\"odinger picture) and
its hermitean conjugate were introduced by \be \varphi({\rm x}):=
\varphi_i({\rm x})a^i, \qquad\quad \varphi^*({\rm x})\!=\!
\varphi_i^*({\rm x}) a^+_i, \qquad\qquad\mbox{(infinite sum over
$i$),} \label{Schfield} \ee where $\{ e_i\}_{i\!\in\!\b{N}}$ is an
orthonormal basis of the 1-particle Hilbert space and $\varphi_i$,
$a^+_i,a^i$ the wavefunction, creation, annihilation operators
associated to $e_i$. Here we summarize how to perform \cite{Fio08}
second quantization on a non-commutative space using a {\it twist}
\cite{Dri83} to deform in a coordinated way space(time), its
symmetries and all objects transforming under space(time)
transformations. This is an alternative to the other approaches to
QFT adopted so far, namely path-integral approaches \cite{Fil96}, or operator
approaches where the field commutation relations are fixed
by other prescriptions, e.g. adapted canonical quantization
(see e.g. \cite{DopFreRob95,AscLizVit07}) 
or Wightman axioms (see e.g. \cite{AlvVaz03,ChaPreTur05,FioWes07});
moreover, as in \cite{Oec00,ChaKulNisTur04,ChaPreTur05,AkoBalJos08,FioWes07}, 
it aims at recovering all the undeformed spacetime symmetries in terms of noncocommutative Hopf algebras .

 A rather general way to
deform an algebra $\A$ (over $\b{C}$, say) into a new one $\A_\star$
is by deformation quantization \cite{BayFlaFroLicSte}. Calling
$\lambda$ the deformation parameter, this means that the two have
the same vector space over $\b{C}[[\lambda]]$,
$V(\A_\star)=V(\A)[[\lambda]]$, but the product $\star$ of
$\A_\star$ is a deformation of the product $\cdot$ of $\A$. On the
algebra $\X$ of smooth functions on a manifold $X$, and on the
algebra $\D\supset\X$ of  differential operators on $\X$,
$f\star h$  can be defined applying to $f\ot h$ first a suitable
bi-pseudodifferential operator $\bF$ (depending on the deformation
parameter $\lambda$ and reducing to the identity when $\lambda=0$)
and then the pointwise multiplication $\cdot$. The simplest example
is probably provided by the Moyal $\star$-product on $X=\b{R}^m$:
\be 
\ba{l} a(x)\!\star\! b(x)\!:=\! a(x) \exp\left[\frac
i2\stackrel{\longleftarrow}{\partial_h}\!\lambda\vartheta^{hk}\!
\stackrel{\longrightarrow}{\partial_k}\right] b(x)=
\cdot\left[\bF(\trc\!\ot\!\trc)(a\ot b)\right],
\\[8pt] \bF
:= \mbox{exp}\left(-\frac i2\theta^{hk}P_{h}\ot
P_{k}\right),\qquad\qquad \theta^{hk}:=\lambda\vartheta^{hk},
\label{Moyalstarprod}\ea 
\ee 
where $P_h$ are the generators of translations (on $\X$
$P_h$ can be identified with $-i\partial_h:=
i\partial/\partial x^h$), and
 $\vartheta^{hk}$ is a fixed real
antisymmetric matrix (this is well-defined for polynomials or other
infinitely differentiable functions $a,b$ such that the
$\lambda$-power series (\ref{Moyalstarprod})$_1$ converges; an
alternative definition (\ref{IntForm}) in terms of  Fourier transforms
 makes sense on much larger domains).

If one replaces
all $\cdot$ by $\star$'s in an equation of motion, e.g. in the
Schr\"odinger equation on $X=\b{R}^3$ of a particle with electrical charge $q$
\be
\ba{c} \Haus\psi({\rm x})=i\hbar  \partial_t
\psi({\rm x}),\quad \qquad
\Haus:=\big[\!\frac{-\hbar^2}{2m}D^a\!\star\!
D_a\!+\!V\big]\star,\quad \qquad D_a\!=\!\partial_a\!+\! i qA_a,
\ea                  \label{1Schr}
\ee
one obtains a pseudodifferential equation and therefore introduces a moderate 
(very special) amount of non-locality in the interactions. In the case of the Moyal
$\star$-product on $X=\b{R}^3$ this becomes
$$
\ba{l} \frac{-\hbar^2}{2m}\partial_h\partial^h\psi({\rm x})\!+\!
V({\rm x})\exp\left[\frac
i2\stackrel{\longleftarrow}{\partial_h}\!\theta^{hk}\!
\stackrel{\longrightarrow}{\partial_k}\right]\psi({\rm
x})=E\psi({\rm x})\ea
$$
where we have chosen for simplicity $A=0$. The kinetic part is
undeformed, non-locality is concentrated only in the interactions. 
$\X_\star,\X$ have the same Poincar\'e-Birkhoff-Witt series, i.e. the subspaces of
$\star$-polynomials and $\cdot$-polynomials of any fixed degree in
$x^h$  coincide. The algebras $\X,\X_\star$ can be defined through
the same generators (i.e. coordinates ${\rm x}^h$ and $\1$) and different
(but related) relations. One can define a linear map
$\wedge:f\!\in\!\X[[\lambda]]\!\to\!\hat f\!\in\!\X_\star$ by the
requirement that it reduces to the identity on the vector space
$V(\X_\star)=V(\X)[[\lambda]]$: $\hat f({\rm x}\star)=f({\rm x})$. One finds
 \be \ba{l}
 \wedge(x^h)=x^h\\
\wedge(x^hx^k)=x^h\star x^k-\frac i2\theta^{hk}
\qquad\Rightarrow\qquad [x^{h}\stackrel{\star}, x^{k}]=\1 i\theta^{hk}\quad\\
\qquad ...\label{defhatx} \ea \ee
In other words, by $\wedge$ one expresses functions of ${\rm x}^h$ as
functions of ${\rm x}^h\star$.
Similarly one proceeds with  $\D,\D_\star$.
$\wedge$ transforms (\ref{1Schr}) into a $\star$-differential equation of
second order (i.e. of second degree in $\partial_h\star$), what
 may critically simplify the study of the equation:
$$
\ba{l}
\frac{-\hbar^2}{2m}\partial_h\star\partial^h\star\hat\psi({\rm
x}\star)\!+\! \hat V({\rm x}\star)\star\hat\psi({\rm
x}\star)=E\hat\psi({\rm x}\star) ,\ea
$$
How does a $\star$-product transform under a transformation of the Euclidean
group $G$, the symmetry group of $\b{R}^3$, or equivalently under the action of the Universal Enveloping  Algebra (UEA) $U\g$ of the
Lie algebra $\g$ of $G$? According to the coproduct of a noncocommutative
Hopf algebra obtained deforming $U\g$ by the twist $\F$ inverse of
(\ref{Moyalstarprod})$_2$ (section \ref{se:2}) . 

Actually, in section \ref{se:2} we are going to  present a procedure
which applies to a large class 
of twist-induced $\star$-deformations of $X=\b{R}^m$ (or 
of symmetric submanifolds $X$ of $\b{R}^m$) and of the spacetime symmetry 
covariance group $G$ of the quantum theories on $X$, leading to results
generalizing the ones sketched above for the Moyal deformations. 

What about multiparticle systems with a non-local interaction of the
above kind? Their description will be simplified if
we use generators $x^h_j\star$, $\partial_h^j\star$,
$j\!=\!1,...,n$. However we respect Bose/Fermi
statistics, i.e. the rule to compute the number of allowed states of
$n$ identical bosons/fermions. Second quantization will be
simplified if we also use generators $a^i\star,a^+_i\star$,
$i\!\in\!\b{N}$.
In general, we shall expand all products $\cdot$'s in terms of
$\star$-products
$$
f\star g=f \cdot g\!+\!\lambda(\bF^\alpha\trc f)\cdot(\bF\!_\alpha\trc
g)\!+\!O(\lambda^2)\qquad\Rightarrow \qquad f \cdot g=f\star
g\!-\!\lambda(\bF^\alpha\trc f)\star(\bF\!_\alpha\trc
g)\!+\!O_\star(\lambda^2)
$$
in {\it all commutative notions} [wavefuncts $\psi$, diff. operators
(Hamiltonian, etc), $a^i,a^\dagger_i$,..., action of $H$, second
quantization itself] to introduce {\it their noncommutative
analogs}. Then forgetting the $\cdot$'s we end up with a "noncommutative
way" to look at QFT, or a {\it noncommutative space(time) and a
(formal) closed framework for covariant QFT on it}.

\medskip
 A similar strategy has been used by J. Wess \&
collaborators \cite{Wes04,AscDimMeyWes06} to formulate noncommutative 
diffeomorphisms and related notions (metric, connections, tensors etc).

\medskip
Alternatively, if one prefers a minimalistic view one can keep a {\it commutative}
spacetime and use $\star$'s only to introduce peculiar non-local
interactions; then the use of noncommutative coordinates may be seen
just as a help to solve the dynamics.

\medskip
In section \ref{se:1} we describe the twist-induced deformation of a cocommutative 
Hopf $*$-algebra and of its module $*$-algebras, in particular the algebras
of functions and of differential operators on symmetric submanifolds $X$ 
of $\b{R}^m$ and the Heisenberg/Clifford algebra associated to 
bosons/fermions on $X$. In section \ref{se:2} we use these tools to
deform the (non-relativistic) wave-mechanical formulation of a system 
of bosons/fermions on $X$ and the Second Quantization of the latter; 
we also study a charged particle in a constant magnetic field ${\bf B}$
on the Moyal deformation of $\b{R}^3$  as an example of a model where the
use of noncommutative coordinates helps solving the dynamics (\ref{1Schr}).
In section \ref{se:3} we extend the Second Quantization procedure to 
 relativistic free fields on a deformed Minkowski spacetime covariant under the
associated deformed Poincar\'e Group, devoting attention in particular
to the Moyal-Minkowski spaces and the
corresponding twisted Poincar\'e Hopf algebra $\widehat{U\P}$
\cite{ChaKulNisTur04,Wes04,Oec00}. 

We denote as $V(\A)$ the vector space
underlying an algebra $\A$.
We stick to linear spaces and algebras over
$\b{C}$ or the ring $\b{C}[[\lambda]]$ of formal
power series in $\lambda$ with coefficients in $\b{C}$;
then tensor products are to be understood as completed in the
$\lambda$-adic topology.
We shall often change notation: \ \ $\X_\star\leadsto\hat \X$,
$\D_\star\leadsto\hat \D$, $x^h_j\!\star\! \leadsto \hat x^h_j$,
$\partial_h^j\!\star\!\leadsto \hat \partial_h^j$, $a^+_i\!\star\!
\leadsto \hat a^+_i$, etc. For instance, the previous Schr\"odinger
equation in the new notation becomes
$$
\ba{l}
\frac{-\hbar^2}{2m}\hat\partial_h\hat\partial^h\hat\psi(\hat{\rm
x})\!+\! \hat V(\hat{\rm x})\hat\psi(\hat{\rm x})=E\hat\psi(\hat{\rm
x}) .\ea
$$

 \section{Preliminaries}\label{1}

\subsection{Twisting $H\!=\!U\g$ to a noncocommutative Hopf
algebra $\hat H$}

The Universal Enveloping  $*$-Algebra (UEA) $H\!:=\!U\g$ of the 
Lie algebra $\g$ of any Lie group $G$
is a Hopf $*$-algebra. First, we briefly recall what this means. Let
$$
\ba{lll}
\varepsilon(\1)=1,\qquad \quad &\Delta(\1)=\1\ot\1,\qquad \quad & S(\1)=\1,\\[8pt]
\varepsilon(g)=0,\qquad \quad & \Delta(g)=g\ot\1+\1\ot g,\qquad
\quad &  S(g)=-g,\qquad \qquad \mbox{if }g\in\g; \ea
$$
$\varepsilon,\Delta$ are extended to all of $H$ as
$*$-algebra maps, $S$ as  a $*$-antialgebra map: \be \ba{lll}
\varepsilon:H\to\b{C},\quad\qquad  & \varepsilon(ab)=\varepsilon(a)\varepsilon(b),\quad\qquad  & \varepsilon(a^*)=[\varepsilon(a)]^*,\\[8pt]
\Delta:H\to H\ot H,\quad \qquad  & \Delta(ab)=\Delta(a)\Delta(b),
\qquad
 & \Delta(a^*)=[\Delta(a)]^{*\ot *},\\[8pt]
S:H\to H,\quad\qquad  & S(ab)=S(b)S(a),\quad\qquad  &
S\left\{\left[S(a^*)\right]^*\right\}=a. \ea \label{deltaprop} \ee
The extensions of $\varepsilon,\Delta,S$ are unambiguous, as
$\varepsilon(g)=0$,
$\Delta\big([g,g']\big)=\big[\Delta(g),\Delta(g')\big]$,
$S\big([g,g']\big)=\big[S(g'),S(g)\big]$ if $g,g'\in\g$. The maps
$\varepsilon,\Delta,S$ are the abstract operations by which one
constructs the trivial representation, the tensor product of any two
representations and the contragredient of any representation,
respectively. $H\!=\!U\g$ equipped with  $*,\varepsilon,\Delta,S$
is a Hopf $*$-algebra.

\medskip
Second, we deform this Hopf algebra. Let $\hH=H[[\lambda]]$. 
Given a {\it twist} \cite{Dri83} (see also
\cite{Tak90,ChaPre94}), i.e. an element $\F\!\in\! (H\ot
H)[[\lambda]]$ fulfilling \bea &&\F=\1\ot\1+ O(\lambda),\qquad
\qquad (\epsilon\ot\id)\F=
(\id\ot\epsilon)\F=\1,                 \label{twistcond}\\[8pt]
&&(\F\ot\1)[(\Delta\ot\id)(\F)]=(\1\ot\F)[(\id\ot\Delta)(\F)]=:\F^3,
                                           \label{cocycle}
\eea
we shall call $H_s\!\subseteq\!H$ the smallest Hopf
$*$-subalgebra such that $\F\!\in\! (H_s\ot H_s)[[\lambda]]$ and
\be
\sum_I \F^{(1)}_I \ot\F^{(2)}_I \!:=\F,\qquad
\quad \sum_I \bF^{(1)}_I \ot\bF^{(2)}_I \!:=\F^{-1},\qquad
\quad\beta:=\sum_I
\F^{(1)}_I S\left(\F^{(2)}_I \right)\in H_s.    \label{defbeta}
\ee
Without loss of generality
 $\lambda$ can be assumed real; for our purposes $\F$ is {\it unitary} ($\F^{*\ot *}=\F^{-1}$),
implying that also $\beta$ is ($\beta^*=\beta^{-1}$). Extending the
product, $*,\Delta,\varepsilon,S$ linearly to the formal power
series in $\lambda$ and setting \be \hat\Delta(g) :=\F \Delta(g)
\F^{-1}, \qquad  \hat S(g):=\beta \, S(g)\beta^{-1} ,\qquad
\R:=\F_{21}\F^{-1},\label{inter-2}
 \ee
one finds that the analogs of conditions (\ref{deltaprop}) are
satisfied and therefore $(\hH,*,\hat\Delta , \varepsilon,\hat S)$ is
a Hopf $*$-algebra deformation of the initial one and has a unitary
triangular structure $\R$ (i.e. $\R^{-1}\!=\!\R_{\scriptscriptstyle 21}\!=\!\R^{*\ot *}$).
While $H$ is cocommutative, i.e.
$\tau\!\circ\!\Delta(g)\!=\!\Delta(g)$ where $\tau$ is the
flip operator [$\tau(a\!\ot\!b)\!=\!b\!\ot\! a$], $\hH$ is
triangular noncocommutative i.e.
$\tau\!\circ\!\hat\Delta(g)\!=\!\R\Delta(g)\R^{-1}$. Correspondingly,
$\hat\Delta , \hat S$ replace $\Delta, S$ in the construction of the tensor product of any two
representations and the contragredient of any representation,
respectively.
Drinfel'd has shown \cite{Dri83} that any triangular deformation of
the Hopf algebra $H$ can be obtained in this way (up to
isomorphisms).

Eq. (\ref{cocycle}), (\ref{inter-2}) imply the
generalized intertwining relation
$\hat\Delta^{(n)}(g)\!=\!\F^n\Delta^{(n)}(g)(\F^n)^{-1}$
 for the iterated coproduct. By definition
$$
\hat\Delta^{(n)}: \hH\to \hH^{\ot n},\qquad\Delta^{(n)}: H[[\lambda]]\to
(H)^{\ot n}[[\lambda]],\qquad\F^n\in (H_s)^{\ot n}[[\lambda]]
$$
reduce to $\hat\Delta,\Delta,\F$ for $n=2$, whereas for $n>2$ they
can be defined recursively as \be \ba{l}
\hat\Delta^{(n\!+\!1)}=(\id^{\ot^{n\!-\!1}}\ot\hat\Delta)
\circ\hat\Delta^{(n)},\qquad\Delta^{(n\!+\!1)}=
(\id^{\ot{(n\!-\!1)}}\ot\Delta)\circ\Delta^{(n)},\\[8pt]
\F^{n\!+\!1}=(\1^{\ot{(n\!-\!1)}}\ot\F)[(\id^{\ot{(n\!-\!1)}}
\ot\Delta)\F^n].                \label{iter-n} \ea \ee
The result for
$\hat\Delta^{(n)},\F^n$ is the same if in definitions (\ref{iter-n})
we iterate the coproduct on a different sequence of tensor factors
[coassociativity of $\hat\Delta$; this follows from the coassociativity of
$\Delta$ and the cocycle condition (\ref{cocycle})];
for instance, for $n\!=\!3$ this amounts to (\ref{cocycle}) and
$\hat\Delta^{(3)}\!=\!(\hat\Delta\ot\id)\!\circ\!\hat\Delta$. For
any $g\in H[[h]]= \hH$ we shall use the Sweedler notations
$$
\Delta^{(n)}(g)=\sum_I  g^I_{(1)} \otimes g^I_{(2)} \otimes ...
\otimes g^I_{(n)},\qquad\qquad
\hat\Delta^{(n)}(g)=\sum_I  g^I_{(\hat 1)} \otimes g^I_{(\hat 2)} \otimes ...
\otimes g^I_{(\hat n)}.
$$

\medskip  For the Euclidean or Poincar\'e group, calling $P_\mu,M_{\mu\nu}$ respectively the generators of translations, homogenous transformations and adopting the Moyal twist (\ref{Moyalstarprod})$_2$ one finds $\beta=\1$, $\hat S= S$ and \bea &&\hat\Delta
(P_\mu)= P_\mu\ot\1+\1\ot P_\mu=\Delta (P_\mu),\nn &&\hat\Delta
(M_\omega)=M_\omega\ot\1+\1\ot M_\omega+
P_\mu([\omega,\theta])^{\mu\nu}\ot P_\nu\neq\Delta
(M_\omega).\nonumber \eea where
$M_\omega\!=\!\omega^{\mu\nu}M_{\mu\nu}$. The Hopf subalgebra of translations is undeformed!

\subsection{Twisting $H$-module $*$-algebras}

We recall that a $*$-algebra $\A$ over $\b{C}$ is defined to be a  left $H$-module
$*$-algebra if there exists a $\b{C}$-bilinear map $(g,a)\!\in\!
H\!\times\!\A\to g\trc a\!\in\!\A$, called (left) action, such that
\be(gg')\!\trc\! a=g \!\trc\! (g'\! \!\trc\! a)\!,\qquad (g\!\trc\!
a)^*\!=[S(g)]^*\!\trc\! a^*\!, \qquad g\!\trc\! (ab)=\!\sum_I\!
\left(g^I_{(1)}\!\trc\! a\right)\!
 \left(g^I_{(2)}\!\trc\! b\right).\qquad        \label{leibniz}
\ee Given such an $\A$, let
$V\!\big(\A\big)$ the vector space underlying $\A$.
$V\big(\A\big)[[\lambda]]$ gets a $\hH$-module $*$-algebra
$\A_\star$ when endowed with the product and $*$-structure
\be\qquad \qquad a\star a':=\sum_I \left(\bF^{(1)}_I \trc a\right)
\left(\bF^{(2)}_I \trc a'\right),\qquad\qquad  
a^{*_\star}:=S(\beta)\trc a^*. \label{starprod} \ee
In fact, $\star$ is associative by (\ref{cocycle}),  fulfills $(a\!\star\!
a')^{*_\star}\!=\!a'{}^{*_\star}\!\!\star\! a^{*_\star}$ and
{}\be \ba{l} g\trc (a\!\star\!a')\!\!=\!\!\sum_I\! \left[g^I_{(\hat
1)}\!\!\trc  a\right] \!\star\! \left[g^I _{(\hat 2)}\!\!\trc
a'\right]. \ea \ee
This is mostly used to deform abelian $\!\A\!$, but works even if $\!\A\!$ is
non-abelian.

Note that the {\it $\star$ is ineffective if $a$ or $a'$ is $H^s$-invariant:}
\be
g\trc a=\epsilon(g) a\quad \mbox{or}\quad g\trc a'=\epsilon(g)
a'\quad \forall g\in H^s \qquad\quad \Rightarrow\qquad \quad a\star
a'=aa'.\qquad \label{Trivstar} \ee

Given $H$-module $*$-algebras $\A,\B$, also $\A\!\ot\!\B$ is,
so (\ref{starprod}) makes $V(\A\!\ot\!\B)$
into a $\hH$-module $*$-algebra $(\A\!\ot\!\B)_\star$. Defining a bilinear map
$\ot_\star$  by
$a\!\ot_\star\! b\!:=\!(a\!\ot\!\1_{\scriptscriptstyle \B})\!\star\!
(\1_{\scriptscriptstyle {\cal A}}\!\ot\! b)$ one finds
\be \ba{l} (a\ot_\star b)\star (a'\ot_\star
b')=\sum_I a\star (\R^{(2)}_I\trc a') \ot_\star (\R^{(1)}_I\trc b)
\star b', \label{braid} \ea\ee 
so $\ot_\star$ is the {\it (involutive) braided tensor product} associated
 to $\R$, and $(\A\!\ot\!\B)_\star=\A_\star\ot_\star\B_\star$.

\medskip
{\bf If $\A$ is defined by generators $a_i$ and relations}, 
{\bf then also $\A_\star$ is}, with the same
Poincar\'e-Birkhoff-Witt series \cite{Fio08}. One can define a {\it linear map} \
\ $\wedge:f\!\in\!\A\!\to\!\hat f\!\in\!\A_\star$ \ \ by the equation
 \be f(a_1,a_2,...)\star=\hat
f(a_1\star,a_2\star,...) \qquad\quad\mbox{in }
V(\A)=V(\A_\star).\label{defhat} \ee This is the generalization of
(\ref{defhatx}). We shall often change notation and replace: \ \
$a_i\!\star\! a_j\leadsto \hat a_i\hat a_j$, $\hat
f(a_i\!\star\!)\leadsto \hat f(\hat a_i)$, $\A_\star\leadsto\hA$, $*_\star\leadsto\hat *$ etc.

\subsection{Application to differential and integral calculi}

 If $\X$ is the  algebra of smooth functions on a manifold
$X$ and $\Xi$ the Lie algebra of smooth vector fields on $X$, one
finds that $\X$ and $\D\!=\!U\Xi\cross \X$ (the algebra of smooth
differential operators on $X$) are $U\Xi$-module $*$-algebras. Each
twist $\F\!\in\! (U\Xi\ot U\Xi)[[\lambda]]$ generates
a$\star$-deformation \cite{AscDimMeyWes06}
$$
\X\stackrel{(\ref{starprod})}{\longrightarrow}\X_\star\sim\widehat{\X},
\qquad\qquad\qquad\D
\stackrel{(\ref{starprod})}{\longrightarrow}\D_\star\sim\widehat{\D}.
$$
Assuming $X$ is Riemannian, let $G$ be its group of isometries,
$H=U\g$, $d\nu$ the $G$-invariant volume form on $X$. Fixed a
$\F\!\in\!(H\ot H)[[\lambda]]$, the invariance of integration [i.e.
$\int_X\!\! d\nu (g\trc f)\!=\!\epsilon(g)\!\!\int_X\!\! d\nu
f$] implies for the corresponding $\star$-product \be \int_X\!\!\!
d\nu({\rm x}) f({\rm x})\star h({\rm x})= \int_X\!\!\! d\nu({\rm x})
f({\rm x}) [\beta^{-1}\trc h({\rm x})] =\int_X\!\!\! d\nu({\rm x})
[S(\beta^{-1})\trc f({\rm x})] h({\rm x}). \qquad
\label{Intstarprop} \ee 
The invariance of the Laplacian $\nabla^2$ implies $\nabla^2\star=\nabla^2$
by (\ref{Trivstar}); moreover, $\nabla^2$ itself can be expressed as a 
$\star$-product of two vector fields, e.g. on $X\!=\!\b{R}^m$ 
$\nabla^2=\partial_h\star\partial_h'$, where 
$\partial_h'\!=\!S(\beta)\trc\partial_h$. As a result, the kinetic part
of the Hamitlonian (\ref{1Schr}) remains undeformed.
For the Moyal $\star$-product on $X\!=\!\b{R}^m$
it is $\beta=\1$, whence $\int_X\!\! d\nu f\star h=\int_X\!\!
d\nu f h$ and $\partial_h'\!=\!\partial_h$.  

\medskip
We now further assume that $X$ admits global coordinates ${\rm x}^h$
(so that $\X$ is generated by the ${\rm x}^h$) and that the map
$\wedge:f\!\in\!\X[[\lambda]]\!\to\!\hat f\!\in\!\X_\star$ is
well-defined (so that $\X_\star$ is generated by the ${\rm
x}^h\star$). This is the case e.g. if $X\subseteq \b{R}^m$ is an
algebraic manifold symmetric under a subgroup $G\subseteq IGL(m)$,
and $\F\!\in\!(H\ot H)[[\lambda]]$, with $H=U\g$ and $\g=Lie(G)$.
Then one can define also a $\hH$-invariant ``integration over $\hat
X$'' $\int_{\hat X}\!\! d\hat\nu(\hat {\rm x})$ such that for each
$f\!\in\!\X$ 
\be 
\int_{\hat X}\!\!\! d\hat\nu(\hat {\rm x}) \hat
f(\hat {\rm x}) =\int_X\!\!\! d\nu({\rm x}) f({\rm x}). 
\ee
We shall call $\wedge^n$ the analogous maps 
$\wedge^n:f\!\in\!\X^{\ot n}[[\lambda]]\!\to\!\hat f\!\in\!(\X^{\ot n})_\star$. 
The previous two equations generalize to integration over $n$
independent ${\rm x}$-variables.

\subsection{Application to the Heisenberg/Clifford algebra $\A^\pm$}

The covariance under a Lie group $G$
of a quantum theory describing a species of
bosons (resp. fermions) implies that the associated Heisenberg
algebra $\A^+$ (resp. Clifford algebra $\A^-$) is a $U\g$-module
$*$-algebra (e.g., for non-relativistic quantum mechanics on
$\b{R}^3$ $G$ is the Euclidean group or its extension the Galilei
group, for the relativistic theory on Minkowski space $G$ is the
 Poincar\'e group). In this subsection we first recall how this happens, then
$\star$-deform $\A^\pm$; we describe the quantum system abstractly
(i.e. in terms of bra, kets, abstract operators). As the Lie group
(of {\it active} transformations) $G$ is unitarily implemented on
the Hilbert space of the system, the action of $H=U\g$ will be
defined on a dense subspace, in particular on a pre-Hilbert space
$\H$ of the one-particle sector, on which it will be denoted as
$\rho$: \
 $g \trc=: \rho(g)\!\in\!\O\!:=\!\mbox{End}(\H)$ \
\ $\forall \!g\!\in\! H$.

The pre-Hilbert space of $n$ bosons (resp. fermions) is described by
the completely symmetrized (resp. antisymmetrized) tensor product
$\H^{\ot n}_+$ (resp. $\H^{\ot n}_-$), which is a $H$-$*$-submodule
of $\H^{\ot n}$.  Denoting as $\vert 0\ra$ the vacuum state, the
bosonic (resp. fermionic) Fock space is defined as the closure
$\bH^{\infty}_{\pm}$ of
$$
\H^{\infty}_{\pm}:=\left\{\mbox{finite sequences
}(s_0,s_1,s_2,...) \in\b{C}\vert 0\ra\oplus\H\oplus
\H^{2}_{\pm}\oplus...\right\}
$$
(finite
means that there exists an integer $l\!\ge\! 0$ such
that $s_n=0$ for all $n\!\ge\! l$). As usual for any orthonormal
basis $\{ e_i\}_{i\in\b{N}}$ of $\H$ we can define an associated set
of creation, annihilation operators for bosons, fermions fulfilling
the Canonical (anti)Commutation Relations (CCR) 
\be
[a^i,a^j]_{\mp}=0,\qquad\quad
[a^+_i,a^+_j]_{\mp}=0,\qquad\quad[a^i,a^+_j]_{\mp} =\delta^i_j\1.
\label{ccr} \ee 
The $H$-invariance of the vacuum $\vert 0\ra$ implies that
creation and annihilation operators $a^+_i,a^i$ must transform as
the vectors $e_i=a^+_i\vert 0\ra$ and $\la e_i,\cdot\ra=\la 0\vert
a^i$ respectively: \be g\trc e_i=\rho_i^j(g)e_j\qquad
\Rightarrow\qquad g\trc a^+_i=\rho_i^j(g) a^+_j,\qquad g\trc
a^i=\rho^\vee{}_i^j(g)a^j = \rho^i_j\big[S(g)\big] a^j
                                       \label{lineartransf}
\ee ($\rho^\vee\!=\!\rho^T\!\circ\! S$ is the contragredient
 of $\rho$). Therefore $\A^{\pm}$ is a $H$-module $*$-algebra because the $\g$-action
(extended to products as a derivation) is compatible with the
(\ref{ccr}).

\bigskip
Applying the deformation procedure one obtains $\hH$-module
$*_\star$-algebras $\A^{\pm}_\star$. The generators $a^+_i$,
$a^{\prime i}\!:=\!a^+_i{}^{*_\star}\!=\!\rho^i_j(\beta)a^j$  fulfill
the $\star$-commutation relations  \be\ba{l}
 a^{\prime i} \!\star\! a^{\prime j} =
\pm R^{ij}_{vu} a^{\prime u}\!\star\!  a^{\prime v} ,\\[8pt]
 a^+_i \!\star\! a^+_j = \pm R_{ij}^{vu} a^+_u \!\star\! a^+_v,\\[8pt]
 a^{\prime i}\!\star\!  a^+_j     = \delta^i_j\1_{\scriptscriptstyle{\cal A}}
\pm R^{ui}_{jv} a^+_u \!\star\! a^{\prime v}, \ea \quad
\Leftrightarrow\qquad  \ba{l} \hat a^{\prime i}\hat a^{\prime j} =
\pm R^{ij}_{vu}\hat a^{\prime u}\hat
a^{\prime v} ,\\[8pt]
\hat a^+_i\hat a^+_j = \pm R_{ij}^{vu}\hat a^+_u\hat a^+_v,\\[8pt]
\hat a^{\prime i}\hat a^+_j     = \delta^i_j\1_{\scriptscriptstyle
\hat{\cal A}} \pm R^{ui}_{jv}\hat a^+_u\hat a^{\prime v}, \ea
\label{hqccr} 
\ee 
where $R\!:=\!(\rho\!\ot\! \rho)(\R)$. The $a^{\prime i}$ transform 
according to the rule of the {\it twisted} contragredient representaton:  
$g\trc a^{\prime i}\!=\! \rho^i_j\big[\hat S(g)\big] a^{\prime j}$. Equivalently,
$\hA^\pm\sim\A^\pm_\star$ has generators $\hat a^+_i,\hat a^i$
fulfilling $\hat a^+_i{}^{\hat *}=\hat a^{\prime i}$ and the
rhs(\ref{hqccr}). Such a general class of equivariantly deformed
Heisenberg/Clifford algebras was introduced in Ref. \cite{Fio96}. Up
to normalization of $R$ the relations at rhs(\ref{hqccr}) are
actually identical to the ones defining the older $q$-deformed
Heisenberg algebras of \cite{PusWor89,Pus89,WesZum90}, based on a
quasitriangular $\R$ in (only) the {\it fundamental} representation
of $H=U_qsu(N)$ (i.e.
 $i,j,u,v\!\in\!\{1,...,N\}$).

\medskip What are the $*$-representations of $\hA^\pm$? Is there a Fock
type one? Yes, on the {\it undeformed} Fock space of
bosons/fermions. The important consequence is that {\bf
(\ref{hqccr}) are compatible with Bose/Fermi statistics}
\cite{FioSch96}.
The simplest explanation of this is that one can "realize" $\hat
a^+_i,\hat a^{\prime i}$ as  "dressed" elements $\check a^+_i,\check
a^{\prime i}$ in $\A^\pm[[\lambda]]$ fulfilling (\ref{hqccr}) and
hermitean conjugate to each other \cite{Fio96}: \be\ba{l} \check
a^+_i=\sum_I \big(\bF^{(1)}_I \!\trc
a^+_i\big)\,\sigma\big(\bF^{(2)}_I \big),\qquad\qquad \check
a^{\prime i}=\sum_I \big(\bF^{(1)}_I \!\trc a^{\prime
i}\big)\,\sigma\big(\bF^{(2)}_I \big). \label{defDf} \ea\ee  In
(\ref{defDf}) we have used the $*$-algebra map
 $\sigma\!:\!H[[\lambda]]\!\rightarrow\! \A^\pm[[\lambda]]$, which is defined by
setting on the generators $\sigma(\!\1_{\scriptscriptstyle
H}\!)\!=\!\1_{\scriptscriptstyle{\cal A}}$, $\sigma(g)\!=\!(g\trc
a^+_j) a^j$ if  $g\in \g$; another characterizing property is that
$$g\trc a=\sum_I \sigma\big(g_{(1)}^I\big)\, a\,\sigma\big(g_{(2)}^I \big)
\qquad\qquad\forall g\in H,\quad a\in \A^\pm.$$

\medskip For $\g=su(2)$ $\sigma$ is the well-known Jordan-Schwinger
realization of $Usu(2)$. For Moyal deformation of $\b{R}^m$, with
generalized basis $\{e_{\rm p}\}$, $P^he_{\rm p}=p^he_{\rm p}$,
(\ref{defDf}) reduces to
$$
\ba{l} \check a^+_{\rm p}=a^+_{\rm p}e^{-\frac i2 p\theta\sigma(P)},
\qquad \qquad \check a^{\rm p}=a^{\rm p}e^{\frac i2
p\theta\sigma(P)},\qquad\qquad \sigma(P^h):=\int \!\!d^m\! p\: p^h
a^+_{\rm p} a^{\rm p}.\ea
$$ 
The latter formulae have already
appeared in the literature (see \cite{Fio08} for a list of references).
Provided the $\lambda$-power series entailed in (\ref{defDf})
converge, $\check a^+_i,\check a^{\prime i}$ are well-defined
operators on the Fock space, providing on the latter also a
representation of $\hA^\pm$. One can also show \cite{Fio08} that
this is the only representation of $\hA^\pm$ of Fock type.

 \section{Non-relativistic second quantization}\label{2}

\subsection{Twisting quantum mechanics in configuration space}

Dealing with a wave-mechanical description of a system of quantum
particles means that the state vectors $s$'s are described by
wavefunctions $\psi$'s on $X$ and the abstract operators by
differential or more generally integral operators on the $\psi$'s.
For simplicity we choose $X=\b{R}^3$, consider spinless
particles and derive consequences from the
covariance of the description first under the Euclidean group $G$
(thought as a group of {\it active}  space-symmetry transformations), then
under the whole Galilei group $G'$. Going to the infinitesimal form,
all elements $H=U\g$ will be well-defined differential operators e.g. on the 
pre-Hilbert space ${\cal S}(\b{R}^3)$, so we can choose $\X$ 
as a dense subspace $\X\subseteq {\cal S}(\b{R}^3)$ (to be specified later)
and tailor the 1-particle pre-Hilbert space $\H$ 
and the algebra of endomorphisms
$\O\!:=\!\mbox{End}(\H)$ as respectively
isomorphic to $\X,\E\!:=\!\mbox{End}(\X)$, by definition;
we shall call the isomorhisms $\kappa,\tilde\kappa$. 
[One reason why we do not identify $\H$ with $\X$ is that we wish to introduce a 
{\it realization} (i.e. representation) of the {\it same} state $s\in\H$ of the quantum system also by a noncommutative wavefunction.]
Summarizing, there exists a (frame-dependent) {\it $H=U\g$-equivariant configuration space realization of $\{\H,\O\}$ on 
$\{\X,\E\}$}, i.e.

\begin{enumerate}\item
there exists a $H$-equivariant, unitary transformation
$\kappa:s\!\in\!\H\leftrightarrow\psi_s\!\in\!\X$, \be g\trc
\psi_s\!=\!\psi_{g\trc s},\qquad\quad \la s |v \ra=\! \int_X\!\!\!
d\nu\: [\psi_s({\rm x})]^*\psi_v({\rm x}).
\label{scalprod1} \ee

\item $\kappa(Os)=\tilde\kappa(O)\kappa(s)$ for any
$s\!\in\!\H$ defines a $H$-equivariant map
$\tilde\kappa\!:\!O\!\in\!\O\!\leftrightarrow\!
D_O\!\in\!\E$, and $\D\subset\E$.

\end{enumerate}

This implies for a system of $n$ distinct particles (resp. $n$ bosons/fermions) 
on $X$:

\begin{enumerate}\item $\kappa^{\ot n}\!:\!\H^{\ot n}\!\leftrightarrow\!\X^{\ot n}$
(resp. the restrictions $\kappa^{\ot n}\!:\!\H^{\ot
n}_{\pm}\!\leftrightarrow\!\X^{\ot n}_{\pm}$) are $H$-equivariant
unitary transformations.

\item  $\tilde\kappa^{\ot n}\!:\!\O^{\ot
n}\!\leftrightarrow\! \E^{\ot n}$ (resp. the restriction
$\tilde\kappa^{\ot n}\!:\!\O^{\ot n}_+\!\leftrightarrow\! \E^{\ot
n}_+$) are $H$-equivariant maps.
\end{enumerate}

\medskip
For any twist $\F\!\in\!(H\ot H)[[\lambda]]$ and the associated
$\star$-deformation one finds 
\be \ba{lll}
\la s, v\ra &=&
\int_X\! d\nu({\rm x}_1)...\!\!\int_X\! d\nu({\rm x}_n) [\psi_s({\rm
x}_1,...,{\rm x}_n)]^*
\psi_v({\rm x}_1,...,{\rm x}_n)\\[8pt]
&\stackrel{(\ref{Intstarprop}), (\ref{starprod})}{=}& \int_X\! d\nu({\rm x}_1)
...\!\!\int_X\! d\nu({\rm x}_n) [\psi_s({\rm x}_1,...,{\rm x}_n)]^{*_\star}
\star\psi_v({\rm x}_1,...,{\rm x}_n). \ea
\ee
If in addition $\F$ is such that 
one can define the map $\wedge$ (see section \ref{se:2}.3), 
we introduce noncommutative
wavefunctions $\hat\psi\!=\!\wedge^n(\psi)$.
Then the previous equation becomes \be \la s, v\ra =\int_{\hat X}\!
d\hat\nu(\hat{\rm x}_1)...\!\!\int_{\hat X}\! d\hat\nu(\hat{\rm
x}_n) [\hat\psi_s(\hat{\rm x}_1,...,\hat{\rm x}_n)]^{\hat
*}\hat\psi_v(\hat{\rm x}_1,...,\hat{\rm x}_n).
\label{scalprodn} \ee The map $\wedge^n:\psi_s\!\in\!\X^{\otimes
n}\!\to\!\hat\psi_s\!\in\!(\X^{\otimes n})_\star$ is thus unitary
and $\hH$-equivariant.

The action of the symmetric group $S_n$ on $(\X^{\otimes n})_\star$
is obtained by ``pull-back'' from that on $\X^{\otimes n}$: a
permutation $\tau\!\in\!S_n$ represented on $\X^{\otimes
n}\!,\!(\X^{\otimes n})_\star$ resp. by the permutation operator
$\P_\tau$ and the ``twisted permutation operator''
$\P^F_\tau\!=\!\wedge^n \P_\tau [\wedge^n]^{-1}$.
Thus, $(\X^{\ot n}_{\pm})_\star,(\E^{\ot
n}_+)_\star$ are
(anti)symmetric  up to the similarity transformation $\wedge^n$ (cf.
\cite{FioSch96}).

\medskip
Let $\hat \kappa^n:=\wedge^n \kappa^{\ot n}$,
$\widehat{\tilde\kappa}^n(\cdot):=\wedge^n [\tilde\kappa^{\ot n}(\cdot)][\wedge^n]^{-1}$.
The restrictions $\hat \kappa^n\!\!\upharpoonright\!_{\!\H^{\ot
n}_{\scriptscriptstyle\pm}}$,
$\widehat{\tilde\kappa}^n\!\!\upharpoonright\!_{\!\O^{\ot
n}_{\scriptscriptstyle +}}$ define a (frame-dependent) {\it
$\hH$-equivariant, noncommutative configuration space realization}
of $\{\H^{\ot n}_{\pm},\O^{\ot n}_+\}$ on $\{(\X^{\ot
n}_{\pm})_\star,(\E^{\ot n}_+)_\star\}$ (or, changing notation, on
$\{\widehat{\X^{\ot n}_{\pm}},\widehat{\E^{\ot n}_{+}}\}$).

\subsection{Quantum fields in the Schr\"odinger picture}

Given a basis $\{ e_i\}_{i\!\in\!\b{N}}\!\subset\!\H$, let $\varphi_i\!=\!\kappa(e_i)$.
The {\bf nonrelativistic field operator}  and its hermitean
conjugate \be \varphi({\rm x}):= \varphi_i({\rm x})a^i,
\qquad\qquad\qquad\qquad \varphi^*({\rm x})\!=\! \varphi_i^*({\rm
x}) a^+_i \ee (infinite sum over $i$) are operator-valued
distributions fulfilling the commutation relations
 \be [\varphi({\rm
x}),\varphi({\rm y})]_{{\scriptscriptstyle \mp}}=\mbox{h.c.}=0,
\qquad \quad [\varphi({\rm x}),\varphi^*({\rm
y})]_{{\scriptscriptstyle \mp}}= \varphi_i({\rm x})\varphi_i^*({\rm
y})=\delta({\rm x}\!-\!{\rm y}) \label{fccr} \ee ($\mp$
for bosons/fermions). The {\it field
$*$-algebra} $\Phi$ can be defined as the span of the normal ordered
monomials 
\be \varphi^*({\rm x}_1)....\varphi^*({\rm x}_m)
\varphi({\rm x}_{m\!+\!1})...\varphi({\rm x}_n) \label{fieldmon}
\ee
(${\rm x}_1,...,{\rm x}_n$ are independent points). So $\Phi\subset
\Phi^e\!:=\!\A^\pm\!\ot\!\left(\bigotimes_{i=1}^{\infty}\!
\X'\right)$ (here the 1st, 2nd,... tensor factor
$\X'$ is the space of distributions depending on ${\rm x}_1,{\rm
x}_2,\!...$); the dependence of 
(\ref{fieldmon}) on ${\rm x}_{h}$ is trivial for $h> n$. $\Phi^e$ is a huge $H$-module $*$-algebra:
$a^+_i,\varphi_i$ transform as $e_i$, and $a^i,\varphi_i^*$
transform as $\la e_i,\cdot\ra$.
The CCR (\ref{ccr}) of $\A^\pm$ are the only nontrivial commutation 
relations in $\Phi^e$.

The key property is that $\varphi,\varphi^*$ are basis-independent,
i.e. {\bf invariant under the group $U(\infty)$ of unitary
transformations of $\{ e_i\}_{i\!\in\!\b{N}}$}, in particular under
the subgroup $G$ of Euclidean transformations (transformations
of the states $e_i$ obatined by translations or rotations of the
1-particle system), or (in infinitesimal form) 
{\bf under $U\g$}: $g\trc \varphi({\rm
x})=\epsilon(g)\varphi({\rm x})$.

As a consequence, if we apply the $\star$-deformation with the above $\F$
we deform $U\g\to\widehat{U\g}$, $Uu(\infty)\to\widehat{Uu(\infty)}$
and the associated module $*$-algebra
$\Phi^e\stackrel{(\ref{starprod})}{\longrightarrow}\Phi^e_\star\sim\widehat{\Phi^e}$,
but find 
\be
\varphi({\rm x})\!\star\! \omega=\varphi({\rm x}) \omega, \qquad
\omega\!\star\!\varphi({\rm x})=\omega\,\varphi({\rm x}), \qquad \quad
\&\quad  \mbox{h. c.},\qquad
\quad\forall\omega\!\in\!V(\Phi^e)[[\lambda]],
\label{fieldinv} \ee
because of (\ref{Trivstar}). Since $\epsilon(\beta)=1$ and the definition
$a^{\prime i}\!:=\!a^+_i{}^{*_\star}\!=\!S(\beta)\trc a^i$ imply
\be
\varphi({\rm x}) =\varphi_i({\rm x}) \star a^{\prime i},
\qquad\qquad\varphi^*({\rm x})=\varphi^{*_\star}({\rm x}) =a^+_i\star \varphi_i^{*_\star}({\rm x}), \ee
and $\varphi_i({\rm x})\varphi_i^*({\rm y})=\varphi_i({\rm
x})\star\varphi_i^{*_\star}({\rm y})$, in $\Phi^e_\star$  the CCR (\ref{fccr}) become 
\be 
[\varphi({\rm x})\stackrel{\star}{,}\varphi({\rm
y})]_{\mp}=h.c.=0, \qquad \qquad\qquad [\varphi({\rm
x})\stackrel{\star}{,}\varphi^{*_\star}({\rm y})]_{\mp}=\varphi_i({\rm x})\star\varphi_i^{*_\star}({\rm y})\qquad\quad
 \label{qfccr}
\ee (here $[A\!\stackrel{\star}{,}\!B]_{\mp}\!:=\!A\star
B\!\mp\! B\star A$). $\Phi^e_\star$ is a huge
$\widehat{U\g}$-module [and also $\widehat{Uu(\infty)}$-module]
$*$-algebra.

\medskip
The unitary
map $\kappa^n_{\pm}:s\!\in\!\H^{\ot n}_{\pm}\leftrightarrow
\psi_s\!\in\!\X^{\ot n}_{\pm}$ and its inverse can be computed using
the field \be\ba{l}
\psi_s({\rm x}_1,\!...,{\rm x}_n)=\frac 1{\sqrt{n!}}
\left\la\left[\varphi^{ *}({\rm x}_1\!)...
\varphi^{ *}({\rm x}_n)\vert 0\ra\right],s\right\ra,\\[8pt]
s=\frac 1 {\sqrt{n!}}\! \displaystyle\int_X \!\! \!\!d\nu({\rm
x}_1\!)...\!\! \displaystyle\int_X \!\!\!\! d\nu({\rm x}_n)
\varphi^{ *}({\rm x}_1\!)...
\varphi^{*}({\rm x}_n)\vert 0\ra \: \psi_s({\rm
x}_1,\!...,{\rm x}_n). \ea \label{wfket} \ee 
 For $s\!\in\!\H^{\ot n}$,  $\psi_s\!\in\!\X^{\ot n}$
 the rhs of these equations give projections 
$\pi^n_\pm\!:\!\H^{\ot n}\!\to\! \X^{\ot n}_{\pm}$ and
$\Pi^n_\pm\!:\!\X^{\ot n}\!\to\!\H^{\ot n}_{\pm}$.
Analogous properties hold also for the deformed counterparts
(see \cite{Fio08}).

\subsection{Equations of motion and Heisenberg picture}

Assume the $n$-particle wavefunction $\psin$ fulfills the
 Schr\"odinger equation (\ref{1Schr}) if $n\!=\!1$,
and 
\be \ba{l}
i\hbar \frac{\partial}{\partial t}\psin=\Hans\psin,\qquad \quad
\Hans\!:= \sum\limits_{h=1}^{n}\Haus({\rm x}_h,\partial_h,t) +\sum\limits_{h<k}
W(\rho_{hk})\star\ea \label{Schr}
 \ee
if $n\!\ge\! 2$; here the time coordinate $t$  remains ``commuting'', and 
the 2-body interaction $W$ depends only on the (invariant) distance 
\ $\rho_{hk}$ \ between ${\rm x}_h,{\rm x}_k$. $\Hans$ will be
hermitean provided $\Hau$ is and $\beta\!\trc\! \Hau\!=\!\Hau$, 
as we shall assume. In general this is a
$\star$-differential, pseudodifferential equation, preserving the
(anti)symmetry of $\psin$. The Fock space Hamiltonian
$$
\Has\!(\varphi)= \int_X\!\!\!\! d\nu({\rm x}) \varphi^{*_\star}({\rm
x})\star\Haus\! \varphi({\rm x})\!\star  + \!\int_X \!\!
\!\!d\nu({\rm x})\!\!\! \int_X \!\!\!\! d\nu({\rm y})
\varphi^{*_\star}({\rm y})\!\star\!\varphi^{*_\star}({\rm x})\!\star\!W(\rho_{{\rm x}{\rm y}})\!\star\!
\varphi({\rm x})\!\star\!\varphi({\rm y})\star
$$
annihilates the vacuum, commutes with the number-of-particles
operator $\mbox{\bf n}\!:=\!a^+_i\!\star a^i$ and its restriction to
$\H^{\ot n}_{\pm}$ coincides with $\Hans$ up to the unitary
transformation $\tilde\kappa^{\ot n}$. As in the undeformed setting,
 formulating the dynamics
on the Fock space allows to consider also more general Hamiltonians
$\Has$, which do not commute with $\mbox{\bf n}$.


The Heisenberg field $\varphihs({\rm
x},t):= [U(t)]^{*_\star}\varphi({\rm x})U(t)$ fulfills the equal
time commutation relations 
\be \ba{l}[\varphihs({\rm
x},t)\stackrel{\star}{,}\varphihs({\rm y},t)]_{\mp}=h.c.=0, \qquad
\quad [\varphihs({\rm x},t)\stackrel{\star}{,}
\varphihs{}^{*_\star}({\rm y},t)]_{\mp}=\varphi_i({\rm
x})\!\star\!\varphi_i^{*_\star}({\rm y})  \\[8pt]
i\hbar \frac{\partial}{\partial
t}\varphihs= [\varphihs,\Has]\qquad\quad\Rightarrow\qquad\quad i\hbar
\frac{\partial\varphihs}{\partial t}=
\Haus\varphihs\quad\:\:\mbox{if } W\!=\!0.\ea\label{Hfccr}
\ee
Eq. (\ref{Hfccr})$_4$ has the same form as (\ref{1Schr}); as
conventional, we shall call ``second quantization'' the replacement 
$\psi\leadsto\varphihs$. If  $\Haus$ is
$t$-independent, so is $\Has$, then $\Has(\varphihs)=\Has(\varphi)$,
and (\ref{Hfccr}) can be formulated directly in
the Heisenberg picture as equations in the unknown $\varphihs(t)$.
The map
$\hat\kappa^{{\scriptscriptstyle H}}(t)=
\hat\kappa\!\circ\! U(t) :\H\to\X_\star$ is a $t$-dependent $\widehat{U\g}$-equivariant unitary map, giving a $t$-dependent noncommutative
configuration space realization of $\H$. We shall denote $\hat\kappa^{\scriptscriptstyle H}(e_i)=\hat\varphi_i(\hat
x)$.

Provided  $\hat V\!=\!\wedge(V)$,
$\hat A^a\!=\!\wedge(A^a)$ are well-defined, replacing
$\hat V({\rm x}\star,t)=V({\rm x},t)\star$, $\hat {\rm
A}({\rm x}\star,t)={\rm A}({\rm x},t)\star$, $\hat\varphi_i({\rm
x}\star)=\varphi_i({\rm x})\star$ we can reformulate the previous
equations  within $\hat\Phi^e$, $\hat\Phi$ using only $\star$-products,
or equivalently, dropping $\star$-symbols and using only
``hatted'' objects:
\be
\ba{l}\hat\varphi(\hat {\rm x}) =\hat\varphi_i(\hat {\rm x})
\hat a^{\prime i},\qquad \qquad\qquad\qquad\qquad\:\:\varphi^{\hat
*}(\hat {\rm
x}) =\hat a^+_i \hat \varphi_i^{\hat *}(\hat {\rm x}) \\[10pt]
[\hat \varphi(\hat {\rm x}),\hat \varphi(\hat {\rm
y})]_{\mp}=\mbox{h.c.}=0, \qquad \qquad\qquad\quad [\hat \varphi(\hat
{\rm x}),\hat \varphi^{\hat
*}(\hat {\rm y})]_{\mp}=\hat\varphi_i(\hat{\rm
x})\hat\varphi_i^{\hat *}(\hat{\rm y}),\\[10pt]
i\hbar \frac{\partial}{\partial t}\hat\psi^{(n)}=\hat\Ha^{(n)}\hat\psi^{(n)},
\qquad\qquad\qquad\qquad\quad\hat \Ha^{(n)}\!=\! 
\sum\limits_{h=1}^{n}\!\hat\Ha^{(1)}
\!(\hat{\rm
x}_h,\hat\partial_h,t)\!+\!\sum\limits_{h<k}\! \hat W(\hat\rho_{hk}),\\[12pt]
\hat\Ha=\!\displaystyle\int_{\hat X} \!\!\!\! d\hat\nu(\hat {\rm x})
\hat \varphi^{\hat *}(\hat {\rm x})\hat\Ha^{(1)}\!(\hat{\rm
x},t)\hat\varphi(\hat{\rm x})  + \!\!\int_{\hat X} \!\! \!\! d\hat\nu(\hat
{\rm x})\!\! \int_{\hat X} \!\!\!\! d\hat\nu(\hat {\rm y})
\hat\varphi^{\hat *}\!(\hat {\rm y})\hat\varphi^{\hat *}\!(\hat {\rm x})
W(\hat\rho_{{\rm x}{\rm y}}) \hat\varphi({\rm x})\hat\varphi({\rm y}),\\[12pt]
[\hat\varphi^{\scriptscriptstyle H}(\hat{\rm
x},t),\hat\varphi^{\scriptscriptstyle H}(\hat{\rm
y},t)]_{\mp}=\mbox{h.c.}=0, \qquad \qquad
[\hat\varphi^{\scriptscriptstyle H}(\hat{\rm
x},t),\hat\varphi^{\scriptscriptstyle H\hat
*}(\hat{\rm y},t)]_{\mp}= \hat\varphi_i(\hat{\rm
x})\hat\varphi_i^{\hat *}(\hat{\rm y}), \\[10pt]
i\hbar \frac{\partial}{\partial t}\hat\varphi_{\scriptscriptstyle
H}= [\hat\varphi^{\scriptscriptstyle H},\hat\Ha],\qquad\qquad\Rightarrow\qquad\qquad i\hbar \frac{\partial}{\partial t}\hat\varphi^{\scriptscriptstyle H}=
\hat\Ha^{(1)}\hat\varphi^{\scriptscriptstyle H}\quad\:\:\mbox{if }
W\!=\!0\ea\label{hatqfccr} \ee
Summing up: at least formally, we can formulate the same theory both
on the commutative and on the noncommutative space, as summarized in (\ref{hatqfccr}).
Solving the dynamics on one or the other will be a matter of
convenience. In a minimalistic view, the noncommutative  setting should be 
adopted only if the
$\hat{\rm x}$-dependence of $\hat V(\hat{\rm x},t),\hat {\rm A}(\hat{\rm
x},t),\hat\varphi_i(\hat{\rm x})$  is simpler than the ${\rm
x}$-dependence  of $V({\rm x},t)$, ${\rm A}({\rm x},t)$,
$\varphi_i({\rm x})$, as it happens e.g. if the latter fulfill
$\star$-differential equations.

In a non-conservative view, this 
construction suggests (\ref{hatqfccr}) as a general 
candidate framework for a {\bf covariant nonrelativistic field quantization on the noncommutative spacetime
$\b{R}\!\times\!\hat X$ compatible with the axioms of quantum mechanics,
including Bose/Fermi
statistics}. Note in particular that
 in both the Schr\"odinger and Heisenberg
picture the (anti)commutator of fields is a ``$c$-number'' distribution.
The framework is not only $\widehat{U\g}$-, but also
$\widehat{U\g'}$-covariant, where $\widehat{U\g'}$ is the Hopf
$*$-algebra obtained from $U\g'$ by the same twist; to account for
the $t$-dependence $C^1(\b{R},\H),C^1(\b{R},\X)$,... must replace
$\H,\X$,... as carrier spaces of the representations. 
Its consistency beyond the level of formal $\lambda$-power series has
to be investigated case by case.  

\subsection{Nonrelativistic Quantum Mechanics on Moyal space 
$\b{R}^3_\theta$: charged particle in a constant magnetic field}

The definition (\ref{starprod})+(\ref{Moyalstarprod})$_2$ for the 
Moyal deformation of the space of smooth functions of $n$ copies
of $\b{R}^m$ (briefly $\b{R}^m_\theta$) can be extended to larger 
domains in terms of Fourier 
transforms. For instance, 
 \be
a(x_i)\star b(x_j)= \int\! d^mh\!\int\!  d^mk\:\, e^{i\left(h\cdot
x_i+k\cdot x_j-\frac{h\theta k}2\right)}\tilde a(h) \tilde b(k).
                                                       \label{IntForm}
\ee
Here $\tilde a(k_i),\tilde b(k_j)$ are the Fourier 
transforms of $a(x_i), b(x_j)$, where
$i,j=1,...,n$ and $x_i\equiv(x_i^h)$, $h=1,...,m$.
The definition of  $\wedge^n$ can be extended unambigously from the space of polynomials in $x_i^h$ to a linear map
$\wedge^n\!:\!\X^{\prime\,\ot n}\!\to\! \widehat{\X^{\prime\,\ot
n}}$, simply by replacing
$x_i\!\to\!\hat x_i\!$ in the Fourier decompositions:
\bea
&&\wedge^n\!\left[\int\!\!\! d^mq_1...\!\!\int\!\!\!  d^mq_n\:\,
e^{iq_1\cdot x_1} \!...e^{iq_n\cdot x_n}\tilde
a(q_1\!,\!...,q_n)\right]\nn
&& :=\!\int\!\!\! d^mq_1...\!\!\int\!\!\!
d^mq_n\:\, e^{iq_1\cdot \hat x_1}\!...e^{iq_n\cdot \hat x_n}e^{\frac i2\sum_{i< j} q_i\theta q_j}\tilde
a(q_1\!,\!...,q_n).                        \label{wedgen}
\eea
For $n=1$ (\ref{wedgen}) is nothing but the well-known Weyl transformation.

It is instructive to see  explicitly how $\wedge^n$, acting on (anti)symmetric
wavefunctions, ``hides'' their (anti)symmetry. Sticking to $n\!=\!2$, we find 
on the basis of plane waves
$$
\wedge^2\left(e^{iq_1\cdot x_1} e^{iq_2\cdot x_2} \pm e^{iq_2\cdot x_1} e^{iq_1\cdot x_2}\right)=
e^{iq_1\cdot \hat x_1} e^{iq_2\cdot \hat x_2}e^{\frac i2 q_1\theta q_2}
\pm e^{iq_2\cdot \hat x_1} e^{iq_1\cdot \hat x_2}e^{-\frac i2 q_1\theta q_2}.
$$
The (anti)symmetry remains
manifest \cite{FioWes07,Fio08} if we use coordinates  $\xi^a_i,X^a$ 
($\xi^a_i\!:=\! x^a_{i\!+\!1}\!-\! x^a_i$, and 
$X^a\!:=\!\sum_{i=1}^nx_i^a/n$ are the coordinates of
the center-of-mass of the system, which are completely symmetric).
The map $\wedge^n$ deforms only the $X$ part
of the wavefunction, leaving unchanged and completely
(anti)symmetric the $\xi$-part. The previous equation e.g. becomes
$$
\wedge^2\left[e^{i(q_1\!+\!q_2)\cdot X} \left(e^{i(q_2\!-\!q_1)\cdot \xi_1} \pm
e^{-i(q_2\!-\!q_1)\cdot \xi_1}\right)\right] =e^{i(q_1\!+\!q_2)\cdot \hat X}
\left(e^{i(q_2\!-\!q_1)\cdot \hat\xi_1} \pm
e^{-i(q_2\!-\!q_1)\cdot \hat\xi_1}\right)
$$

\medskip
As an example of a simple 1-particle model where the
use of noncommutative coordinates helps solving the dynamics (\ref{1Schr}) 
we now consider
a charged particle on $\b{R}^3_\theta$ in a constant magnetic field ${\rm B}$. 
The simplest gauge choice is $A^i(x)\!=\!\epsilon^{ijk}B^j x^k/2$.
One finds that $\Haus$ is still differential of second order, but more
complicated than its undeformed (i.e. $\theta\!=\!0$) counterpart. 
In terms of ``hatted'' objects the model can be formulated and
solved as in the undeformed case. We choose the $x^3$-axis parallel to
$q{\rm B}=qB\vec{k}$ with $qB>0$, this gives $\hat
D^3\!=\!\partial^3$, $\hat D^a\!=\!\partial^a\!-\!i\frac{qB}{2\hbar
c}\epsilon^{ab}\hat x^b$ for $a,b\!\in\!\{1,2\}$, with
$\epsilon^{12}\!=\!1\!=\!-\epsilon^{21}$, $\epsilon^{aa}\!=\!0$.
These fulfill $[\partial^3,\hat D^a]\!=\!0$, $[\hat D^1,\hat
D^2]\!=\!i\frac{qB}{\hbar c}[1\!-\!\frac{qB\theta^{12}}{2\hbar c}]$.
Defining 
\bea 
\ba{l} a\!:=\!\alpha[\hat D^1\!-\!i\hat D^2],\qquad
a^*\!=\!\alpha[-\hat D^1\!-\!i\hat D^2]\qquad
\alpha\!:=\!\sqrt{\frac {\hbar
c}{qB}}/\sqrt{2\!-\!\frac{qB\theta^{12}}{2\hbar c}}\ea 
\eea (we
assume $qB\theta^{12}\!<\!4\hbar c$) one obtains the commutation
relation $[a,a^*]=1$, and 
\be\ba{l}
\hat\Hau\!=\!\frac{-\hbar^2}{2m}\hat D^i\! \hat D^i
\!=\!\frac{-\hbar^2}{2m}\left[(\partial^3)^2\!-\!\frac 1{2\alpha^2}(aa^*+a^*a)\right]\!=\!\hat\Hau_\parallel\!+\!\hat\Hau_\perp\\[8pt]
\hat\Hau_\parallel\!:=\!\frac{(-i\hbar\partial^3)^2}{2m},\qquad
\hat\Hau_\perp\!:=\!\hbar\omega\left(a^*a\!+\!\frac 12\right),\qquad
\omega\!:=\!\frac {qB}{mc}\left(1\!-\!\frac{qB\theta^{12}}{4\hbar
c}\right) \ea\ee  
It is easily checked that $[\hat\Hau_\parallel,\hat\Hau_\perp]=0$.
$\hat\Hau_\parallel$ has continuous spectrum $[0\!,\!\infty[$; the
generalized eigenfuntions are the eigenfuntions $e^{ik\hat x^3}$ of
$p^3=-i\hbar\partial^3$ with eigenvalue $\hbar k$. The second is
formally an harmonic oscillator Hamiltonian with $\omega$ modified
by the presence of the noncommutativity $\theta^{12}$.
So the spectrum of $\hat\Hau$
is the set of  $E_{n,k^3}\!=\!\hbar\omega (n\!+\!1/2)+(\hbar k^3)^2/2m $.

To find a basis of eigenfunctions we define in analogy with the undeformed case
$$
\ba{l} \hat z\!:=\!\sqrt{\!\frac {\zeta}2}(\hat x^1\!\!+\!i\hat
x^2),\quad
\partial_{\hat z}\!:=\!\frac1{\sqrt{2\zeta}} (\partial_1\!-\!i\partial_2),\qquad\Rightarrow\qquad
\hat z^*\!=\!\sqrt{\!\frac {\zeta}2}(\hat x^1\!-\!i\hat x^2),
\quad\partial_{\hat z}^* \!=\!-\partial_{\hat z^*} \ea
$$
with $\zeta\!:=\!qB/2\hbar c$, and find that the only nontrivial
commutators among $\hat z,\hat z^*,
\partial_{\hat z},\partial_{\hat z^*}$ are
$$
\ba{l} [\partial_{\hat z},\hat z]=1,\qquad [\partial_{\hat z^*},\hat
z^*]=1,\qquad[\hat z,\hat z^*]\!=\!\zeta\theta^{12}. \ea
$$
We can thus re-express $a,a^*$ in the form
$$
a= \alpha\sqrt{2\zeta}(\hat z^*\!+\!\partial_{\hat z}), \qquad \quad
a^*= \alpha\sqrt{2\zeta} (\hat z\!-\!\partial_{\hat z^*}).
$$
Setting $\hat l^3\!:=\!\hat z\partial_{\hat z}\!-\!\hat
z^*\partial_{\hat z^*} \!-\!\zeta\theta^{12}\partial_{\hat
z}\partial_{\hat z^*}$ and ${\rm n}\!:=\!a^*a$ we also find
\be
[\hat l^3,\hat z^*]=-\hat z^*
,\qquad [ \hat l^3,\hat z]=\hat z,
\qquad [\hat l^3,a^*]= a^*,\qquad [ \hat l^3,a]=- a,
\qquad [ \hat l^3,{\rm n}]=0.
\ee
In analogy with the undeformed case we can therefore choose as a complete 
set of commuting observables $\{p^3,{\rm n} ,\hat l^3\}$. Let
\be
\hat\psi_{0,0}(\hat z^*,\hat z)\!:=\!\int\!\!dk\,dk^*e^{ik\hat z^*}
e^{ik^*\hat z} e^{-kk^*} \qquad\Rightarrow\qquad
a\hat\psi_{0,0}=0=\hat l^3\hat\psi_{0,0}.
\ee
(when $\theta=0$ this gives $\psi_{0,0}\propto e^{-zz^*}$).
The {\it deformed Landau eigenfunctions} 
\be 
\hat\psi_{k^3,n,m}(\hat {\rm
x})=(a^*)^n(\hat z^*)^{n\!-\!m}\hat\psi_{0,0} (\hat z^*,\hat
z)e^{ik^3\hat x^3}
\label{DefLandauEigenf}
\ee 
are generalized eigenfunctions with eigenvalues $p^3\!=\!\hbar
k^3\!\in\!\b{R}$, $,{\rm n}\!=\!n\!=\!0,1,...$, $\hat
l^3\!=\!m\!=n,n\!-\!1,...$ and build up an orthogonal basis of
$\L^2(\b{R}^3)$. 
They are also eigenfunctions of $\hat\Hau$ with
eigenvalues $E_{n,k^3}\!=\!\hbar\omega (n\!+\!1/2)+(\hbar k^3)^2/2m
$. Each energy level, in particular the lowest one, 
has $\infty$-ly many different eigenfunctions, which are characterized by
the same $n,k^3$ and different $m$'s.
Replacing $\hat x^a\!\to\! x^a\star$ and performing all the $\star$-products
one finds the corresponding eigenfuntions
$\psi_{k^3,n,m}({\rm x})$ of $\Haus$. Their ${\rm x}$-dependence is messy,  
whereas the $\hat{\rm x}$-dependence of the (\ref{DefLandauEigenf}) is simple and
practically the same as in the undeformed counterpart, as anticipated.

\section{Relativistic second quantization}
\label{3}

By analogous considerations one can construct (at least) a
consistent free QFT on a noncommutative 
Minkowski spacetime with twisted symmetry.
The commutative manifold $X$ is
Minkowski spacetime with coordinates 
$x\!\equiv\!(x^\mu)$ ($\mu=0,1,2,3$ and $x^0\!\equiv\! t$)  w.r.t. a fixed
inertial frame, and $G$ is its symmetry group, the Poincar\'e Lie group. 
A relativistic particle is described choosing as the
algebra of observables $\O=H=U\g$ and as the Hilbert space
$\bH$ the completion of a pre-Hilbert space $\H$ carrying an
irreducible $*$-representation of $U\g$ charaterized by a
nonnegative eigenvalue $m^2$ of the Casimir $P^\mu P_\mu$ and a
nonnegative spectrum for $P^0$. We stick to the case
of a scalar particle of positive mass $m$. 
 
The relevant space of functions is 
$\X\!:=\!\kappa^{\scriptscriptstyle H}(\H)\!\subset\!{\cal
S}(\b{R}^4)$, the pre-Hilbert space of rapidly decreasing, smooth,
positive-energy solutions of the Klein-Gordon equation. 
The map $\kappa^{\scriptscriptstyle H}$ is the usual $U\g$-equivariant ($t$-dependent) commutative configuration space realization of $\H$.
We denote $\kappa^{\scriptscriptstyle H}(e_i)=\varphi_i$ (these functions depend both on space and time)
and $\Phi^e=\A^{\pm}\!\ot\!\left(\bigotimes_{i=1}^{\infty}\! \X'\right)$,
where $\X'$ stands for the dual space of $\X$.
Fixed a twist $\F\!\in\!(H\ot H)[[\lambda]]$ (e.g. the Moyal twist,
or the twists classified in \cite{BorLukTol}) we apply the associated
$\star$-deformation procedure $H\to\hH,\Phi^e\to\Phi^e_\star$, etc.
The hermitean relativistic free field (in the Heisenberg picture) is
expressed in the form
\be
\varphi(x)=\varphi_i(x)\!\star\!a^{\prime i}
\!+\!a_i^+\!\star\!\varphi_i^{*_\star}(x)=\!\! 
\int \!\!\!\frac{d^3p}{2p^0} [e^{-ip\cdot x}\!\star\! a^{\prime {\rm p}}
\!+\!a_{\rm p}^+ \!\star\! e^{ip\cdot x}\,]  \label{fielddeco} 
\ee
and the free field commutation relation in the form
\be
[\varphi(x)\stackrel{\star},\varphi(y)]=
\varphi_i(x)\star\varphi_i^{*_\star}(y)\!-\!\varphi_i(y)\star
\varphi_i^{*_\star}(x)= \int\!
\frac{d^3p}{2p^0}\left[e^{-ip\!\cdot \!(x\!-\!y)}\!-\!e^{ip\!\cdot
\!(x\!-\!y)}\right]
\ee
for any $x,y\!\in\!\{x_1,x_2,...\}$. This vanishes if
$x\!-\!y$ is space-like (microcausality). We have expressed the right-hand sides
both in terms of a generic normalizable basis 
$\kappa^{\scriptscriptstyle H}(e_i)=\varphi_i$ of $\X$ and in terms
of the generalized basis 
$\kappa^{\scriptscriptstyle H}(e_{{\rm p}})\!=\! e^{-ip\cdot
x}$ of eigenvectors of the momentum operators
$\tilde\kappa^{\scriptscriptstyle H}(P_\mu)\!=\!i\partial_\mu$ with eigenvalues
$p_\mu$ [$(p^a)\!\equiv\!{\rm p}\!\in\!\b{R}^3$ and
$p^0\!\equiv\!\sqrt{{\rm p}^2+m^2}\!>\!0$]; $a^{\rm p},a_{\rm p}^+$ are the associated generalized creation \& annihilation operators, which fulfill
$$
[a^{\rm p},a^{\rm q}]=0,\qquad\quad [a^+_{\rm
p},a^+_{\rm q}]=0,\qquad\quad [a^{\rm p},a^+_{\rm q}]
=2p^0\delta^3({\rm p}-{\rm q}). \label{ccr'} 
$$  
Assuming that $\F$ is such that $\wedge$ is
well-defined on the whole of $\X$ and going to the ``hat notation''
we find 
\be 
\ba{l} \hat\varphi(\hat x)=\hat\varphi_i(\hat x)\hat
a^{\prime i} \!+\!\hat a_i^+\hat\varphi_i^{\hat
*}(\hat x)\\[8pt]
 [\hat\varphi(\hat x),\hat\varphi(\hat y)]= \hat\varphi_i(\hat
x)\hat\varphi_i^{\hat
*}(\hat y)\!-\!\hat\varphi_i(\hat y)\hat\varphi_i^{\hat
*}(\hat x)\\[8pt]
(\hat\Box+m^2)\hat\varphi(\hat x)=0      \ea \label{freecomm} 
\ee

\medskip
Choosing the Moyal twist (\ref{Moyalstarprod})$_2$ one obtains Moyal-Minkowski
noncommutative spacetime and the twisted Poincar\'e Hopf algebra
of \cite{ChaKulNisTur04,Wes04,Oec00}. 
In terms of generalized creation \& annihilation operators
formulae (\ref{hqccr}), (\ref{defDf}) and the states created by
the deformed creation operators become
 \bea
&&\ba{l}
 a^{+}_{\rm p} \!\star\! a^{+}_{\rm q}= e^{-i
 p\theta q}\,     a^{+}_{\rm q} \!\star\! a^{+}_{\rm p}, \\[8pt]
a^{\rm p}\!\star\!  a^{\rm q}=
e^{-i p\theta q} \,    a^{\rm q} \!\star\!  a^{\rm p}, \\[8pt]
a^{\rm p}\!\star\!  a^{+}_{\rm q}=e^{i p\theta q} \,  a^{+}_{\rm q} \!\star\! a^{\rm p}\!+\!
2p^0\delta^3({\rm p}\!-\!{\rm q})\\[8pt]
a^{\rm p}\!\star\! e^{iq\cdot x} =e^{-ip\theta q} \,e^{iq\cdot x}\!\star\! a^{\rm p},
\quad \& \mbox{ h.c.},
\ea
\qquad \Leftrightarrow\qquad
\ba{l}\hat a^+_{\rm p} \! \hat a^+_{\rm q}= e^{i
q\theta\! p}\,     \hat a^+_{\rm q} \! \hat a^+_{\rm p},\\[8pt]
\hat a^{\rm p}  \hat a^{\rm q}\!=\! e^{i q\theta\! p}     \hat a^{\rm q}
\! \hat a^{\rm p},\\[8pt]
\hat a^{\rm p}\! \hat a^+_{\rm q}\!=\!
e^{i  p\theta\! q}   \hat a^+_{\rm q} \! \hat a^{\rm p}\!+\!2
p^0\delta^3\!({\rm p}\!-\!{\rm q}),\\[8pt]
\hat a^{\rm p} e^{iq\cdot \hat x} =e^{-ip\theta q} \,e^{iq\cdot \hat x} \hat a^{\rm p},\quad \& \mbox{ h.c.} ;
\ea\quad\label{aa+cr}\\[8pt]
&& \check a^+_{\rm p}\equiv D_{\f}^{\sigma}\left(a^+_{\rm p}\right)=a^+_{\rm p}e^{-\frac i2 p\theta\sigma(P)}, \qquad \qquad \check a^{\rm p}\equiv
D_{\f}^{\sigma}\left(a^{\rm p}\right)=a^{\rm p}e^{\frac i2 p\theta\sigma(P)}
\label{defA"}\\
&&\hat a^+_{{\rm p}_1}...\hat  a^+_{{\rm p}_n}\vert 0\ra=
a^+_{{\rm p}_1}\star...\star a^+_{{\rm p}_n}\vert 0\ra=
\check a^+_{{\rm p}_1}...\check a^+_{{\rm p}_n}\vert 0\ra=
\exp\!\left[-\frac i2\!\sum\limits_{\stackrel{j,k=1}{j<k}}^n 
p_j\theta p_k\right]\!
a^+_{{\rm p}_1}...a^+_{{\rm p}_n}\vert 0\ra \label{genstates}
\eea
where the Jordan-Schwinger map takes the form
$\sigma(P_\mu)=\int \!d\mu(p)\,p_\mu  a^+_{\rm p} a^{\rm p}$. By (\ref{genstates})
 generalized states differ from their undeformed counterparts only by multiplication by a  phase factor. 
As $\check a^+_{\rm p}\check a^{\rm p}\!=\!a^+_{\rm p}a^{\rm p}$,
$\sigma(P_\mu)\!=\!\int \!d\mu(p)\,p_\mu \check a^+_{\rm p}\check
a^{\rm p}$, from (\ref{defA"}) the inverse of $D_{\f}^{\sigma}$ is
readily obtained.

It is remarkable that the free field  (\ref{fielddeco}) \&
(\ref{aa+cr}) coincides with the one found in formulae (37) \& (46)
of \cite{FioWes07} [see also formulae (32) \& (36) of
\cite{Fio08Proc}] imposing just the free field equation and Wightman
axioms (modified only by the requirement of {\it twisted} Poincar\'e
covariance). See \cite{FioWes07,Fio08Proc,Fio08} for comparisons
with other approaches considered in the literature. In
\cite{FioWes07} it has been also shown that the $n$-point functions
of a (at least scalar) field theory, when expressed as functions of
coordinates differences $\xi^\mu_i\!:=\! x^\mu_{i\!+\!1}\!-\! x^\mu_i$, 
coincide with the undeformed ones. Quite disappointingly,  this result holds 
in time-ordered perturbation theory also for interacting
fields with interaction $\varphi^{\star n}$, due to the translation
invariance of the latter, but should no more hold for e.g. 
a quantum matter field interacting with a gauge field.

\end{document}